\documentclass[final,5p,times,twocolumn]{elsarticle}

\usepackage{amssymb,amsmath,color,hyperref}

\let\oldref\ref
 \renewcommand{\ref}[1]{(\oldref{#1})}

\begin{document}

\begin{frontmatter}



\title{Species coexistence in a neutral dynamics with environmental noise}


\author{Jorge Hidalgo}
\ead{hidalgo@pd.infn.it}
\author{Samir Suweis}
\author{Amos Maritan}
\address{Dipartimento di Fisica e Astronomia, 'G. Galilei' and CNISM, INFN, Universit\`a di Padova, Via Marzolo 8, 35131 Padova, Italy}

\begin{abstract}
Environmental fluctuations have important consequences in the organization of ecological communities, and understanding how such a variability influences the biodiversity of an ecosystem is a major question in ecology. In this paper, we analyze the case of two species competing for the resources within the framework of the neutral theory in the presence of environmental noise, devoting special attention on how such a variability modulates species fitness. The environment is dichotomous and stochastically alternates between periods favoring one of the species while disfavoring the other one, preserving neutrality on the long term. We study two different scenarios: in the first one species fitness varies linearly with the environment, and in the second one the effective fitness is re-scaled by the total fitness of the individuals competing for the same resource. We find that, in the former case environmental fluctuations always reduce the time of species coexistence, whereas such a time can be enhanced or reduced in the latter case, depending on the correlation time of the environment. This phenomenon can be understood as a direct consequence of Chesson's storage effect.
\end{abstract}


\end{frontmatter}



\section{Introduction}
\label{sec:introduction}
One of the main problems in theoretical biology relies on the search for mechanisms leading to the conservation of biodiversity \citep{Hooper2005}. 
Looking at natural systems, it still remains unclear how some ecosystems are able to maintain such a large variety of species \citep{McCann2000}, such as in tropical forests \citep{Volkov2005}, phytoplankton in oceans \citep{Vargas2015}, and coral reefs \citep{Sale1977}, to name but a few. More generally, explaining the stability of large complex ecological networks remains an open and debated issue \citep{Montoya2006,Allesina2012,Suweis2014} and many works have proposed different mechanism as possible  contributors in the maintenance of biodiversity in both trophic \citep{Virginia2014,Allesina2015} and mutualistic \citep{Bascompte2007,Suweis2015} communities.

Abiotic conditions such as the temperature, light, precipitations, humidity, available nutrients in soil, etc., strongly influence the organization and biodiversity of ecological systems \citep{Dunson1991}. Furthermore, immutable environments could be considered an oddity in Nature \citep{Pearman2008}.
Many theoretical studies have tried to explain the impact of environmental fluctuations on population growth and ecosystem stability  \citep{Lewontin1969, Vergassola2015, May1973, Chevin2010} and its influence on evolutionary dynamics \citep{Levins-book, Frank1990, Ashcroft2014}; others have analyzed the role of environmental changes in prey-predator dynamics \citep{Luo2007,Zhu2009,Dobramysl2013}, dispersal \citep{McPeek1992,Yoshimura1996} and the development of survival mechanisms to deal with unpredictable environments, usually referred to as bet-hedging strategies \citep{Kussell2005}.

The question about how biodiversity can be maintained have been also addressed within the framework of the Neutral Theory of Biodiversity (NTB) \citep{Hubbell2001, Azaele2016}.
This paradigm establishes a perfect equivalence among individuals, and, despite being a simple theory, has been able to describe and understood many ecological patters observed in Nature \citep{Hubbell2001, Azaele2016}. Only recently, some works have studied the impact of environmental noise in neutral dynamics \citep{Nadav2014, Nadav-ecoletters, Nadav-Samir2015, Vergassola2015}. For instance, it has been argued that, although the NTB leads to successful predictions for static patterns, the theory fails to estimate several dynamical measures \citep{Nadav-ecoletters}, such as the scaling of species abundance fluctuations with the total population size. Environmental noise seems to fix these issues while preserving the previously reported phenomenology for the static patterns \citep{Nadav2014, Nadav-ecoletters, Nadav-Samir2015}.

Nevertheless, the role of environmental variability in maintaining the biodiversity of communities of neutral species is an open question. Indeed, it still needs to be clarified whether environmental noise has a positive or negative impact on species coexistence. For instance the authors of reference  \citep{Vergassola2015} study the dynamics of bacterial communities growing under limited conditions that respond differently to environmental fluctuations but are neutral on average. They show that environmental noise always reduces the possibility of species to coexist. In contrast, in another recent work \citep{Nadav2016}, authors analyze the impact of the environment in a time-average neutral metacommunity model, showing that, under certain conditions, the total number of species supported by the ecosystem increases due to the variability of the environment. This can be viewed as a direct consequence the so-called \textit{storage effect} evidenced by Chesson and Warner in 1981 \citep{Chesson1981}. Furthermore, it has been reported that, in order to obtain such a mechanism, it is crucial that environmental stochasticity affects recruitment instead of mortality rates \citep{Chesson1981,Nadav-ecoletters}. However, a deep understanding of this issue from a theoretical point of view is still missing.

The goal of this paper is to shed some light on this variety of phenomenologies. To this end, we focus on the simple scenario in which two species compete for the resources with the dynamics of the voter-model \citep{Castellano2009} in a well-mixed situation (i.e. neglecting spatial effects), with the key ingredient that the rates at which species colonize new sites vary with the environment. The model constitutes a general framework in which different dynamics (e.g. environmental variability affecting species mortality instead of recruitment rates, etc.) are mapped into different functional dependencies of the model parameters on the environmental variables. 
For each scenario, we compute analytically and numerically the mean time of coexistence before one of the species monodominates in the community, and we show that such a time can be enhanced or reduced by the effect of the environment depending on the specific case and on the characteristic time correlation of the environment. We provide a general model that to helps clarify what is the net effect of the environment in neutral communities, but the specific dynamics has to be chosen depending on the particularities of the real system under consideration.

\section{Voter model with environmental noise}
\label{sec:voter}
The voter model was first formulated in the context of social dynamics to study how different opinions ``compete'' in a social network until, eventually, a general consensus is reached \citep{Castellano2009}. Different variants have been devised to analyze ecological problems with great success, in particular the voter model with speciation \citep{Azaele2016} of the Neutral Theory of Biodiversity to which we have already referred. 

Here we analyze the simple case of two competing species without speciation nor migration, and consider a fixed population of $N$ individuals that can be either of species $A$ or $B$. 
For the generalization of the model with environmental variability, it is convenient to introduce different fitnesses $\lambda_A$ and $\lambda_B$ for species $A$ and $B$, respectively. We restrict our analysis to the case of a well-mixed community (i.e. mean-field) in which the spatial organization of the community is not considered.

In the dynamics, one individual is randomly chosen at each time step with uniform probability, removed from the population and replaced by a copy of one of its neighbors (in our case any individual in the community) with a probability proportional to its fitness. This process can be mapped into the following set of ``chemical reactions'':

\begin{equation}
\begin{array}{ccc}
 A + B &\overset{\lambda_A}{\longrightarrow}& A + A\\
 A + B &\overset{\lambda_B}{\longrightarrow}& B + B.
\end{array}
\label{eq:reaction}
\end{equation}
Let us note that the previous formulation is also valid for a dynamics with asymmetric mortality and equal recruitment rates, i.e. when the probability of removing an individual of species A (B) is not uniform but proportional to its mortality rate $d_A$ ($d_B$), and the vacant place is occupied by a copy of a random neighbor with uniform probability. In such a case, Eq. \ref{eq:reaction} still holds if we replace $\lambda_{A,B}\rightarrow d_{B,A}$.

In a neutral scenario, species fitnesses are constant and equal, $\lambda_A=\lambda_B$, and species abundance in the population changes only due to demographic fluctuations.
Eventually, one of the species can monodominate and  the dynamics stops. It is known that, for well-mixed populations, the time to reach such a monodominant state in the voter model scales linearly with the population size \citep{Castellano2009}. This constitutes our point of reference when analyzing the impact of environmental fluctuations on species coexistence.

\begin{figure}[t!]
 \centering\includegraphics[width=\columnwidth]{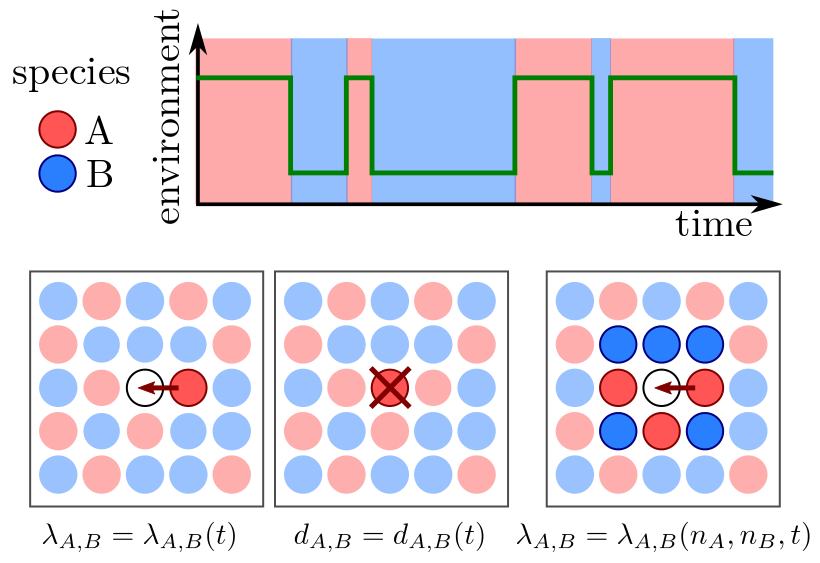}
 \caption{A community of individuals of two species, $A$ and $B$ (identified with colors red and blue, respectively), competing for the available resources in a lattice (sites) with the dynamics of the voter model. \textbf{(Top panel)} Environment changes in time, alternating between periods that favors one of the species (corresponding shaded region) while disfavors the other one. 
 We distinguish several cases depending on how the environment modulates mortality and/or recruitment rates: \textbf{(Bottom Left Panel)} An individual of species A (resp. B) occupies one of its adjacent places at rate $\lambda_A(t)$ (resp. $\lambda_B(t)$), independently of its surrounding neighbors.  \textbf{(Bottom Central Panel)} Equivalently, an individual of species A (similarly for B) is killed at rate $d_A(t)$ (resp. $d_B(t)$), and then replaced by a random neighbor with uniform probability. For these two cases, species fitness does not depend on the local species density. In contrast, in \textbf{(Bottom Right Panel)}, an individual of species $A$ (and similarly for $B$) colonizes one if its adjacent sites, but the colonization rate is re-scaled by the total local fitness. This leads to a more complicated situation in which recruitment depends on the number of neighbors of species $A$ and $B$, that we call $n_A$ and $n_B$, respectively, so that $\lambda_A=\lambda_A(n_A,n_B,t)$.}
 \label{fig:1}
\end{figure}

We aim to model a situation in which species fitness depends on external, variable, conditions. We consider the simple case in which the state of the environment is encoded in a random variable, $\epsilon=\epsilon(t)$, that alternates between two possible states, $\epsilon(t)=\pm1$, at constant rate $k$ (as sketched in top panel of Fig. \ref{fig:1}), i.e. the environment is described by a dichotomous Markov noise (DMN). The choice of DMN stems from several reasons: i) it allows for mathematical treatment (see \citep{Bena2006} for a review on the theory of DMN), ii) it has a finite correlation time, $\tau=(2k)^{-1}$ \citep{Bena2006}, and iii) fluctuations are bounded, in contrast with other colored noises such as the Ornstein-Uhlenbeck process \citep{Gardiner}.
With this choice, species fitness in Eq. \ref{eq:reaction} becomes time dependent, $\lambda_{A,B}\rightarrow\lambda_{A,B}(t)=\lambda_{A,B}(\epsilon(t))$, and there is an additional reaction equation for the environmental variable:

\begin{equation}
\begin{array}{ccc}
 A + B &\overset{\lambda_A(\epsilon)}{\longrightarrow}& A + A\\
 A + B &\overset{\lambda_B(\epsilon)}{\longrightarrow}& B + B\\
 \epsilon &\overset{k}{\longrightarrow}& -\epsilon.
\end{array}
\label{eq:reaction-env}
\end{equation}
Species fitnesses can differ from time to time, but neutrality among species is conserved \textit{on average}, so that $\langle\lambda_A(t)\rangle = \langle\lambda_B(t)\rangle$, where $\langle \cdot \rangle$ refers to the temporal average.

In the well-mixed scenario, the state of the system is fully represented by the number of individuals of species $A$, $n_A$, and the state of the environment, $\epsilon$. 
We can write the Master Equation for the probability of finding the system in a state $(n_A,\epsilon)$, and after performing a Kramers-Moyal expansion in terms of the species A density, $x=n_A/N$, we find the following stochastic equation (see Appendix A for details):
\begin{eqnarray}
 \dot x &=& \big[\lambda_A(\epsilon) - \lambda_B(\epsilon)\big] x (1-x) +\nonumber\\
	& &\dfrac{1}{\sqrt{N}} \sqrt{\big[\lambda_A(\epsilon) + \lambda_B(\epsilon)\big]x(1-x)}\xi(t),
 \label{eq:langevin-voter-env}
\end{eqnarray}
where $\xi(t)$ represents a Gaussian white noise of unit amplitude, $\langle \xi(t)\xi(t') \rangle = \delta(t-t')$, and $\epsilon$ follows its independent dynamics. The last term in the previous equation is due to the demographic noise/fluctuations.
Eq. \ref{eq:langevin-voter-env} has two states in which the dynamics stops, $x=0$ and $x=1$, usually known as ``absorbing'' states in the language of stochastic processes, representing the cases in which species B and A monodominates, respectively.

Before proceeding, it is worth noting some considerations about our approach. We are interested in analyzing the limit of very large populations, so the second term of Eq. \ref{eq:langevin-voter-env} can be safely neglected. 
Such an approximation provides us with an equation with one single source of stochasticity, known as a `piecewise deterministic Markov process', and the theory of DMN applies straightforwardly \citep{Bena2006}. 

However, demographic noise is responsible of taking the population into the absorbing state, and in principle we are not able to compute the time of coexistence before mono-dominance if we neglect the second term in Eq. \ref{eq:langevin-voter-env}. 
We circumvent this problem by identifying the absorbing state with the state in which only one individual of the two species still survives, i.e. with $x=1/N$ or $x=1-1/N$ \citep{Ricardo2012}. As we will show, this simple approximation provides the correct scaling relations obtained from numerical simulations of the individual-based model.
It would be interesting to study a mesoscopic scenario and analyze the interplay between demographic and environmental noise (see for instance \citep{Nadav2016}), and in fact some analytical approaches have been recently devised to frame such a challenging task \citep{McKane2015}. Here we restrict our calculations to the case in which environmental noise is the only source of stochasticity.

In the following section, we present two specific choices for the way in which the fitness depends on the environment, that we call the linear and relative fitness cases, respectively, and that can be mapped into different plausible dynamics in the ecosystem. For each case, we study analytically and numerically the mean time of species coexistence.

\section{Results}
\subsection{Linear fitness case and variability in mortality}
\label{sec:voter-env1}
The simplest way to introduce environmental variability in the dynamics is to assume a linear dependency on environmental fluctuations in species fitness, $\lambda_{A,B}(\epsilon)=c^1_{A,B}+c^2_{A,B}\epsilon$. This choice corresponds to the case in which individuals of species $A$ (and similarly for $B$) colonize neighboring places that are occupied by individuals of species B at --time-dependent-- rate $\lambda_A(t)$, independently of its neighbors' fitness (see left bottom panel of Fig. \ref{fig:1}). Equivalently, this situation can be mapped into the case in which environment variability affects mortality rates instead of species fitnesses, $d_{A,B}(\epsilon)=d^1_{A,B}+d^2_{A,B}\epsilon$ (see central bottom panel of Fig. \ref{fig:1}). In both cases, \textit{no community-level effects enter in the formulation of the fitness}. Focusing on this case, we model a situation in which each time the environment favors one of the species it disfavors the other one:
\begin{equation}
 \lambda_A(\epsilon) = \lambda+\dfrac{\sigma}{2}\epsilon, \qquad \lambda_{B}(\epsilon) = \lambda-\dfrac{\sigma}{2}\epsilon,
 \label{eq:rates-envI}
\end{equation}
where $\lambda$ represents the average fitness and $\sigma$ the variability of the environment (with the constraint that $\sigma\leq2\lambda$). A more general partially-correlated case could be modeled, but this leads to qualitatively similar results, as far as the correlation between $\lambda_A$ and $\lambda_B$ is not one (as discussed in \citep{Vergassola2015}).

Using Eq. \ref{eq:rates-envI}, the dynamics given by Eq. \ref{eq:langevin-voter-env} (neglecting demographic fluctuations) follows the simple equation:

\begin{equation}
\dot x = \sigma\epsilon(t) x(1-x).
\label{eq:langevin-voter-env-I}
\end{equation}
We wish to compute the mean time in which the process described by Eq. \ref{eq:langevin-voter-env-I} reaches any of the (artificial) absorbing boundaries, $x=1/N$ and $x=1-1/N$.
To gain some intuition, it is first convenient to calculate the stationary distribution associated to such a process in the whole interval $[0,1]$ (if there exists), that we call $P_{st}(x)$. The form of $P_{st}(x)$ for a general stochastic process driven by DMN is well known \citep{Bena2006}, and it has been included in Appendix B. Naively, for the process of Eq. \ref{eq:langevin-voter-env-I}, we obtain $P_{st}(x)=C\big(x(1-x)\big)^{-1}$, where $C$ is a normalization constant. However, this distribution cannot be normalized in the interval $[0,1]$ because of the non-integrable divergences at $x=0,1$. These divergences mean that the only stationary solution is the state of mono-dominance \citep{Munoz1998}. More importantly, our results suggest that is precisely the environmental noise that may drive the population towards such a state, reducing the time of species coexistence.

To compute the mean-time in which one of the species dominates, we follow the work of Sancho \citep{Sancho}, who developed a general framework to compute mean-first passage times in stochastic processes driven by DMN (see details in Appendix C). In short, the analytical approach consists in calculating the ``backwards'' equation for the probability distribution $P(x,t)$ with initial condition $P(x,0)=\delta(x-x_0)$, so that the backwards dynamics can be written as $\partial_t P(x,t) = L^\dag_{x_0} P(x,t)$, where $L^\dag_{x_0}$ is the backwards evolution operator. Then, mean first passage times can be found by solving the equation $L^\dag_{x_0} T_N(x_0)=-1$ imposing appropriate boundary conditions. This leads to a second-order differential equation for $T_N(x_0)$ \citep{Sancho} (see Appendix C), that for the process described by Eq. \ref{eq:langevin-voter-env-I} is:

\begin{equation}
 \left(x_0(1-x_0)\right)^2 T_{N}''(x_0) + x_0(1-x_0)(1-2x_0) T_N'(x_0) + \dfrac{1}{\tau\sigma^2} = 0,
 \label{eq:mfpt1}
\end{equation}
with boundary conditions $T_N(1/N) = T_N(1-1/N) = 0$. The solution to this equation is:
\begin{eqnarray}
 T_N(x_0) &=& \dfrac{1}{2\tau\sigma^2}\left(
\log^2\left(N-1\right)
 -
\log^2\left(x_0^{-1}-1\right)
 \right).
 \label{eq:mfpt1-solution}
\end{eqnarray}
In the limit of large $N$ we obtain the simple scaling rule $T_N \simeq  \dfrac{1}{2\tau\sigma^2} (\log N)^2$, in contrast with the linear scaling found for the standard case without environmental noise, $T_N\simeq N$ \citep{Castellano2009}. This result leads to the important conclusion that, if species fitness is modulated by the environment in the way of Eq. \ref{eq:rates-envI}, the time in which species coexist is much reduced by environmental fluctuations. We note that the mean extinction time given by Eq. \ref{eq:mfpt1-solution} diverges for $\sigma\rightarrow0$, however this is merely a consequence of ignoring the demographic noise.

\begin{figure}[t!]
 \centering\includegraphics[width=\columnwidth]{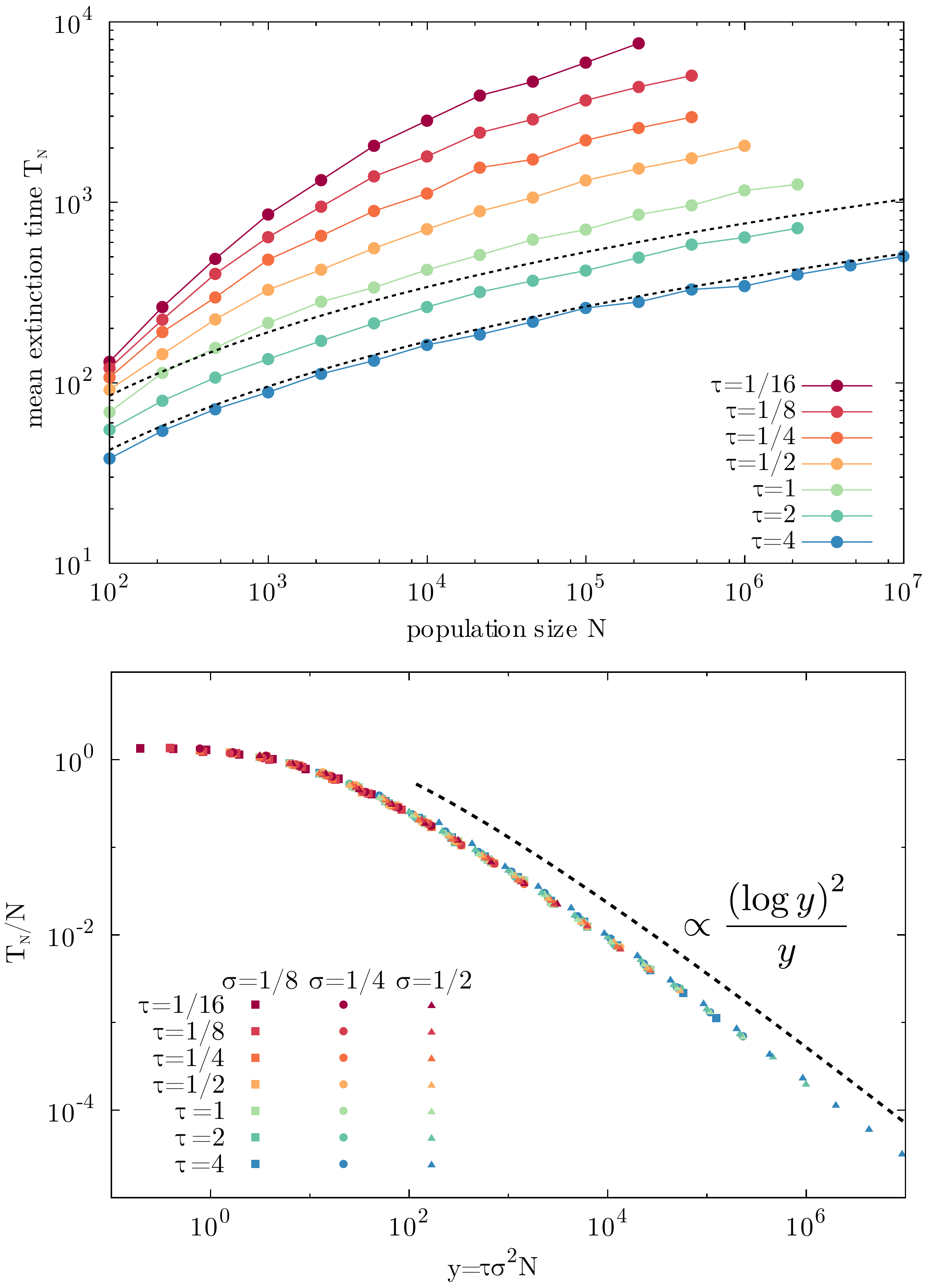}
 \caption{Coexistence in a neutral dynamics in which the species fitness changes linearly with environmental fluctuations (Eq. \ref{eq:rates-envI}).
 {\bf (Top Panel)} Mean time to reach mono-dominance computed with the individual-based model, as a function of the population size $N$, $T_N$, for different values of the environmental correlation $\tau$; environmental variability was fixed to $\sigma=0.25$, but similar curves can be found for different values of $\sigma$. In all cases, environmental fluctuations lead to a sublinear dependency on the population size, in contrast with the linear scaling for the case without environmental fluctuations \citep{Castellano2009}, evidencing the negative effect of the environment for the species coexistence. Dashed lines represent the prediction of $T_N$ (Eq. \ref{eq:mfpt1-solution}) for $\tau=2$ and $\tau=4$, which has been obtained when environmental fluctuations are dominant and demographic noise can be neglected, i.e. for $\tau\sigma^2\gg1/N$. Our analytical prediction is extremely good for $\tau=4$ (and larger), while it gives the correct asymptotic behavior for smaller $\tau$.{\bf (Bottom Panel)} We can collapse the data in a single scaling relationship, $F(y)$, for different values of the population size $N$, correlation $\tau$ and environmental variability $\sigma$, plotting $T_N/N$ as a function of $\tau\sigma^2 N$. $F(y)$ behaves as a constant for small arguments, $y\ll1$, and as $\left(\log y\right)^2/y$ for large arguments, $y\gg1$ (Eq. \ref{eq:scaling1}).}
\label{fig:2}
 \end{figure}

We carried computer simulations of the individual based model, i.e. of the system represented by Eq. \ref{eq:reaction-env}, using Eq. \ref{eq:rates-envI}, by means of the Gillespie algorithm \citep{Gillespie2007}. Starting from a homogeneous initial condition, i.e. $n_A=N/2$ and a random value for the environment $\epsilon=\pm1$ with equal probability, we calculated the mean time to reach mono-dominance for different values of the population size $N$, environmental variability $\sigma$ and correlation $\tau$. Let us note that $\tau$ is given in terms of the mean lifetime of an individual (in the voter model, all the individuals are replaced, on average, each $\Delta t=1$).
Results are plotted in Fig. \ref{fig:2}, illustrating the expected sub-linear scaling with the system size for large $N$.

We can give a step forward and write a heuristic full scaling relation taking into account both demographic and environmental fluctuations.
To do so, we need a parameter that quantifies the ``strength'' of environmental noise. A good candidate is the product $\sigma^2 \tau$, that weighs the variance of environmental fluctuations by its correlation time. If environmental noise is negligible compared to demographic fluctuations ($\sigma^2 \tau\ll 1/N$), we should retrieve the linear relation $T_N\simeq N$ of the standard voter model without environmental noise \citep{Castellano2009}; in the opposite scenario in which environmental fluctuations are dominant ($\sigma^2 \tau \gg 1/N$), we found that $T_N\simeq (\log N)^2/2\tau\sigma^2$. Put together, we can write that $T_N(N)= N F(\sigma^2 \tau N)$, with the scaling function:

\begin{equation}
F(y) = 
\left\{
\begin{array}{ll}
 f_1, & y\rightarrow 0\\
 f_2\dfrac{(\log y)^2}{y}, & y\rightarrow \infty,
\end{array}
\right.
\label{eq:scaling1}
\end{equation}
in which $f_1$ and $f_2$ are constant values. Fig. \ref{fig:2} (bottom panel) represents $T_N/N$ as a function of $\sigma^2 \tau N$, for different values of $N$, $\tau$ and $\sigma$. It can be seen that all the points fall in a perfect collapse, illustrating a functional dependency of the form given by Eq. \ref{eq:scaling1}.

\subsection{Relative fitness case}
\label{sec:voter-env2}
Environmental variability can be introduced in the dynamics in a rather different way, leading to drastic changes in the phenomenology. 
As a guiding example (see bottom right panel of Fig. \ref{fig:1}), let consider a ``forest'' of species $A$ and $B$ in which individuals produce a number of ``seeds'' that varies with environmental conditions in a linear form, 
$s_{A,B}(\epsilon)=c^1_{A,B}+c^2_{A,B}\epsilon$. In the dynamics, one individual is killed at random and the vacant site is colonized by one seed from the neighboring trees. The probability of selecting one of the species is then proportional to the number of seeds of such species \textit{divided by} the total number of seeds produced by the neighbors. This example leads to a density-dependent fitness, which in the well-mixed scenario is:

\begin{equation}
 \lambda_{A,B}(x,\epsilon)= \dfrac{s_{A,B}(\epsilon)}{x s_A(\epsilon) + (1-x) s_B(\epsilon)}.
 \label{eq:lambda-s}
\end{equation}
Note that, in contrast with the linear case, \textit{species fitness becomes density-dependent} through the denominator of Eq. \ref{eq:lambda-s}. A dynamics with individual fitness in the form of Eq. \ref{eq:lambda-s} was first studied by Chesson and Warner \citep{Chesson1981}, reporting the so-called storage effect that we illustrate in what follows.

Eq. \ref{eq:langevin-voter-env} can be generalized straightforwardly for the case in which the rates $\lambda_{A,B}$ also depend on the density $x$ (see Appendix A). As in the previous section, we restrict our calculations to the simple anticorrelated case in which $s_A(\epsilon)=s+\dfrac{\sigma}{2}\epsilon$ and $s_{B}(\epsilon)=s-\dfrac{\sigma}{2}\epsilon$, and without loss of generality, we take $s=1/2$.  Doing this, the stochastic differential equation for the species A density becomes:
\begin{eqnarray}
 \dot x &=& \dfrac{s_A(t)-s_B(t)}{x s_A(t) + (1-x)s_B(t)} x(1-x) \nonumber\\
   &=&2 \dfrac{\sigma\epsilon(t)}{1+\sigma(2x-1)\epsilon(t)} x(1-x).
\end{eqnarray}
Multiplying and dividing by $1-\sigma(2x-1)\epsilon(t)$ (as $\epsilon^2=1$), we get the following linear-in-$\epsilon$ equation:

\begin{equation}
 \dot x = \dfrac{2\sigma x(1-x)}{1-\sigma^2(2x-1)^2}\big( -\sigma(2x-1) + \epsilon(t) \big).
\label{eq:langevin-voter-env-II}
 \end{equation}
As in the previous section, one can obtain some intuition about the time of coexistence by computing the stationary distribution associated to the process described by Eq. \ref{eq:langevin-voter-env-II}. In this case, the solution can be normalized in the range $[0,1]$ (see Appendix B), which is:

\begin{equation}
 P_{st}(x) = \dfrac{\Gamma(1/\tau)}{\Gamma(1/2\tau)^2} \left[x(1-x)\right]^{1/{2\tau}-1},
 \label{eq:pst-dmn}
\end{equation}
where $\Gamma$ represents the Euler Gamma function. Let us note that we can write a non-trivial stationary solution because we have neglected demographic fluctuations, otherwise the only stationary solutions correspond to the absorbing states, $\delta(x)$ and $\delta(1-x)$. Eq. \ref{eq:pst-dmn} can be derived for more general forms of environmental noise in the limit of small variability \citep{Chesson1989} (see also \cite{Nadav2016}).

\begin{figure}[h!]
 \includegraphics[width=\columnwidth]{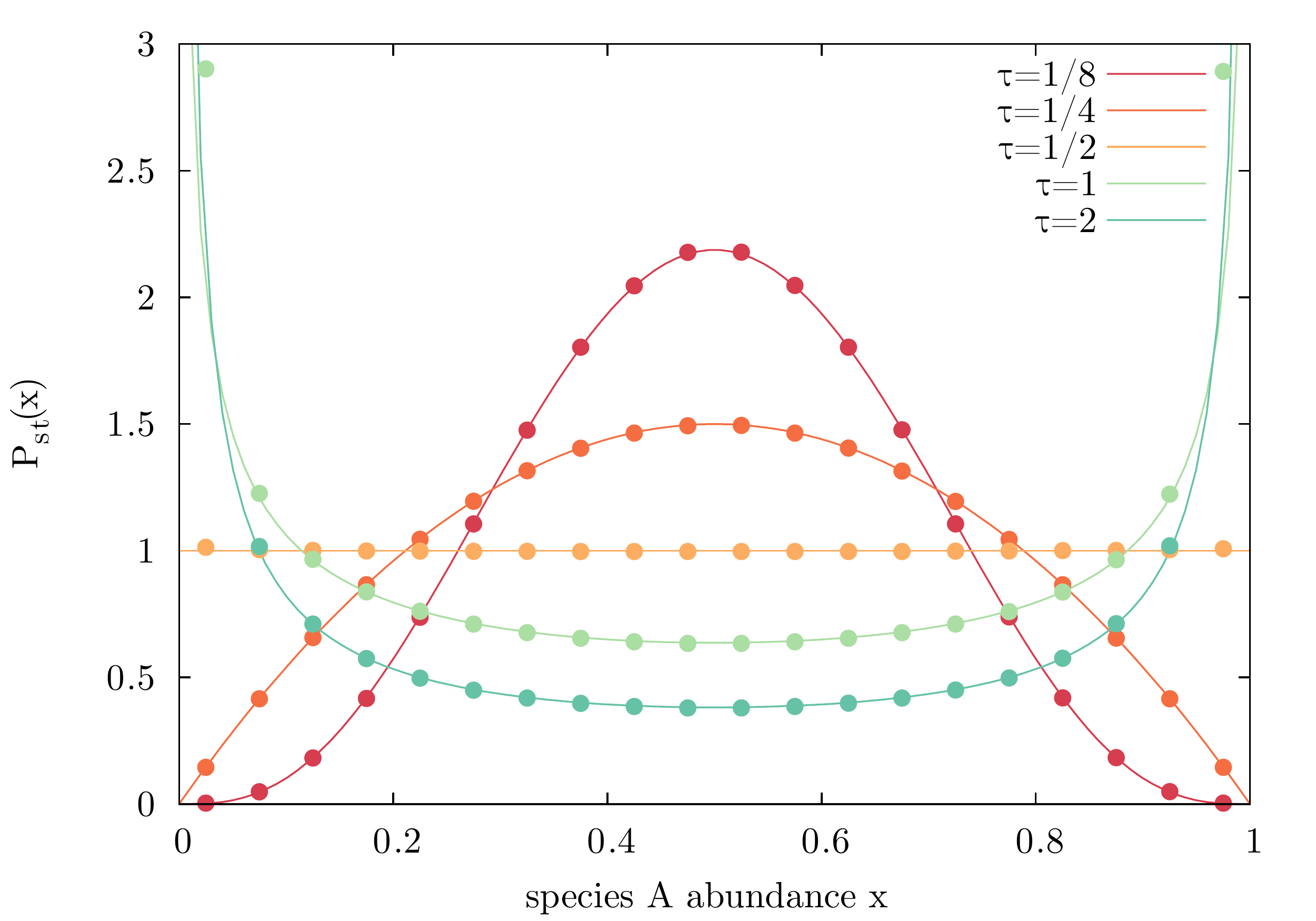}
 \caption{Stationary distribution in the infinite-size community limit in which the species fitness is modulated by Eq. \ref{eq:lambda-s}, for different values of the correlation of the environment, $\tau$.  Continuous lines represent the theoretical prediction (Eq. \ref{eq:pst-dmn}) while dots have been found by integrating numerically Eq. \ref{eq:langevin-voter-env-II} using $\sigma=0.25$ (similar results are obtained for other values of this parameter). We find a distribution peaked at $x=1/2$ for rapidly changing environments ($\tau<1/2$), enhancing species coexistence. On the other hand, slowing changing environments ($\tau>1/2$) give a distribution peaked at the borders, increasing the probability of one species to monodominate. A flat distribution is found for the critical case, $\tau=1/2$. This phenomenon is known as the Chesson's storage effect \citep{Chesson2000}.
}
 \label{fig:3}
\end{figure}

The stationary distribution of Eq. \ref{eq:pst-dmn} is represented in Fig. \ref{fig:3} for different values of the correlation of the environment, evidencing that $\tau$ plays a fundamental role in the shape of $P_{st}(x)$. For rapidly changing environments ($\tau<1/2$), the distribution is peaked at $x=1/2$, i.e. environmental noise enhances species coexistence. Instead, for slowly changing environments ($\tau>1/2$), the distribution is peaked at the borders, in detrimental of species coexistence once demographic fluctuations can push species towards extinction. The stationary distribution is flat for the critical case, $\tau_c=1/2$.
In conclusion, environmental noise may favor or disfavor coexistence depending on the correlation time of the environment.  From this result, we hypothesize (and we confirm later) that the mean time of coexistence scales superlinearly with the population size for $\tau<1/2$, and sublinearly for $\tau>1/2$.

The reported phenomenology is a direct consequence of having a fitness that becomes density-dependent when the environment varies (i.e. for any $\sigma>0$). Indeed, density-dependent fitnesses are well known to shape species coexistence in voter-like dynamics \citep{Borile2012}. Intuitively, Chesson's storage effect results as a byproduct of different responses to the environment depending on the species abundance \citep{Chesson2000}: species with many individuals are highly decimated during unfavorable conditions (relatively to its abundance), while small populations are (proportionately) less damaged and rapidly grow when the environment becomes favorable; consequently an effective drift favors species to coexist. The interplay between such a mechanism and how persistent is the environment makes the temporal correlation a fundamental parameter in the phenomenon.

As discussed in \citep{Nadav2016}, the stationary solution given by Eq. \ref{eq:pst-dmn} does not depend on $\sigma$. This seems to be paradoxical as the impact of the environment is independent on the amplitude of environmental variability. However, the time to reach the stationary state diverges for $\sigma\rightarrow0$. Additionally, demographic fluctuations become more relevant if the variability of the environment is reduced, and the previous derivation may not hold. For a more detailed discussion on the interplay between environmental and demographic noise and Chesson's storage effect we refer to \citep{Nadav2016}.

Finally, we compute the mean-time of coexistence for the dynamics described by Eq. \ref{eq:langevin-voter-env-II}. 
To do so, we can write a similar equation to Eq. \ref{eq:mfpt1}, that can be solved in the limit of $N\gg1$ (see Appendix D), obtaining:
\begin{equation}
 T_N\sim N^{\dfrac{1}{2\tau}}.
 \label{eq:ext2}
\end{equation}
The origin of such a power-law behavior can be elucidated using an approximate formula for the calculation of mean-first passage times, which
can be understood as a modified Arrhenius' law \citep{Sancho} (see Appendix C): introducing the effective Arrhenius factor $U(x)$, the mean time to overcome a barrier located at $x=x^*$ starting from $x=x_0$, up to exponential order, is $T_N(x_0) \sim \exp\left(U(x^*)-U(x_0)\right)$. In the case of Eq. \ref{eq:langevin-voter-env-II}, the Arrhenius factor becomes logarithmic in $x$ (see Appendix C),  $U(x) = -\dfrac{1}{2\tau} \log\left[x(1-x)\right]$ (that, remarkably, is independent of the environmental variability $\sigma$). Then, evaluating at any of the (symmetrical) artificial absorbing barriers ($x=1/N$ or $x=1-1/N$) such a logarithmic potential leads to the power-law scaling of Eq. \ref{eq:ext2}.

Note that, if $\tau<1/2$, the mean time of coexistence increases respect to the case without environmental noise ($T_N\sim N$), favoring species coexistence. On the other hand, when $\tau>1/2$, the time is reduced to a sub-linear dependency, i.e. disfavoring coexistence. For the critical case, $\tau=1/2$, we find the same scaling relation to the case without environmental noise, i.e. linear with the population size $N$. 
In all cases, mean extinction times scale as a power-law, which is a characteristic signature of noise-induced fixed points \citep{Kessler2007,odorico}.

\begin{figure}[t!]
 \includegraphics[width=\columnwidth]{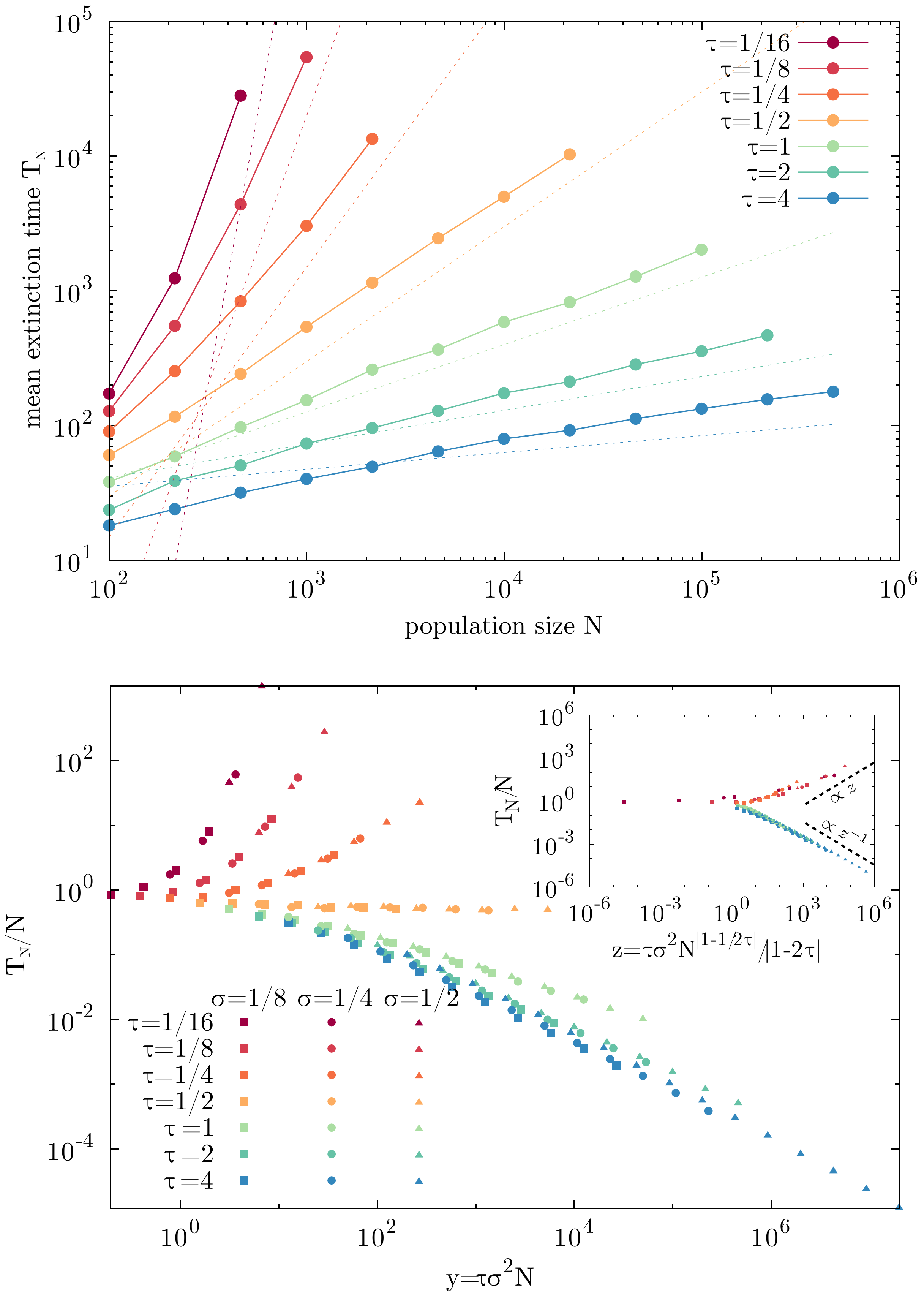}
 \caption{Coexistence in a neutral dynamics in which the species fitness changes with environmental fluctuations, normalized by the overall fitness in the community  (Eq. \ref{eq:lambda-s}).
 {\bf (Top Panel)} Mean time to reach mono-dominance as a function of the population size $N$, computed with a simulation of the individual-based model, for different values of the correlation $\tau$. Environmental variability has been set to $\sigma=0.25$, but, qualitatively, similar curves are found for other values of $\sigma$. Dashed lines represent the asymptotic behavior expected from the theoretical prediction, $T_N\sim N^{1/2\tau}$. In contrast with the results found in Fig. \ref{fig:2}, a super-linear scaling relationship is found for rapidly changing environments $\tau<1/2$, whereas the contrary occurs for slowly changing environments, $\tau>1/2$.
 {\bf (Bottom Panel)} The main plot shows a partial collapse for different values of the population size $N$ and environmental variability $\sigma$, showing different scaling functions for each correlation time $\tau$, as given by Eq. \ref{eq:scaling2}. The inset represents a full approximate collapse of the same data, one for $\tau<1/2$ and another one for $\tau>1/2$, in good agreement with the scaling relation proposed in Eq. \ref{eq:scaling3}.
 }
 \label{fig:4}
\end{figure}

We run computer simulations of the individual-based model (i.e. the system represented in Eq. \ref{eq:reaction-env} with Eq. \ref{eq:lambda-s}) by means of the Gillespie's algorithm \citep{Gillespie2007}, for different values of the population size $N$, environmental correlation $\tau$ and variability $\sigma$. The initial condition was taken as in the previous case, i.e. $n_A=N/2$ and a value for the environment at random with equal probability.
The mean time to reach mono-dominance is plotted in Fig. \ref{fig:4}. We can see a power-law behavior in perfect agreement with Eq. \ref{eq:ext2} for large population sizes. 
Power-laws of variable exponent are commonly found in the presence of environmental noise, that can be related with the so-called ``temporal Griffiths phases'' in the context statistical mechanics \citep{temporal-griffiths}.

We can develop a scaling relationship taking into account both the demographic and the environmental noise. We hypothesize that
$T_N\simeq N G(\tau\sigma^2N, \tau)$, with the scaling function:

\begin{equation}
G(y,\tau) = 
\left\{
\begin{array}{ll}
 g_1 & y\rightarrow 0\\
 g_2(\tau) y^{\dfrac{1}{2\tau}-1} & y\rightarrow \infty,
\end{array}
\right.
\label{eq:scaling2}
\end{equation}
where $g_1$ is a constant and $g_2(\tau)$ is a function of $\tau$, so that $T_N\simeq N$ in the regime controlled by demographic noise ($\tau\sigma^2\ll1/N$) and $T_N\simeq N^{1/2\tau}$ in the regime of environmental noise ($\tau\sigma^2\gg1/N$).
The collapses are represented in Fig. \ref{fig:3}B, illustrating a well-agreement with Eq. \ref{eq:scaling2}.

We also make an ansatz for a full scaling relationship valid for all values of $\tau$.
The best collapse was found introducing the new variable $z=\left(\dfrac{1}{\left|\tau-\dfrac{1}{2}\right|}\tau\sigma^2N\right)^{\left|1-\dfrac{1}{2\tau}\right|}$, so that we can still distinguish the regime controlled by demographic fluctuations, $z\ll1$, and the regime controlled by environmental noise, $z\gg1$. Let us note that this relation is only valid for $\tau\neq1/2$. With this, $T_N = N \bar G_\tau(z)$, with the scaling function:

\begin{equation}
\bar G_\tau(z) = 
\left\{
\begin{array}{ll}
 \bar g_1         & z\rightarrow 0\\
 \bar g_2 z       & \tau<1/2, z\rightarrow \infty\\
 \bar g_2' z^{-1}  & \tau>1/2, z\rightarrow\infty \end{array}
\right.
\label{eq:scaling3}
\end{equation}
where $\bar g_1$ $\bar g_2$ and $\bar g_2'$ are constant values. The result is represented in the inset of Fig. \ref{fig:4}B, illustrating a relatively well agreement with Eq. \ref{eq:scaling3}. Analytical understanding of this scaling relationship would require to go beyond the leading order in Eq. \ref{eq:ext2}, that we think is beyond the scope of this work.

\section{Discussion}
In this work, we have studied a model of two species competing for the available resources with the dynamics of the voter model \citep{Castellano2009, Azaele2016}, with the key ingredient that colonization rates depend on a external dichotomous environment.
Periodically, environmental noise increases the fitness of one species while disfavor the other one. This situation is reverted with a characteristic time scale, that is kept as a control parameter through our analysis, and neutrality among species is preserved on average. In this context, we study if environmental noise increases or reduces the time of coexistence before one species monodominates in the community.

This model constitutes a general framework in which different dynamics can be mapped into different functional dependencies of the fitness in terms of the environmental variable. In particular, we study three situations: \textit{i)} the case in which recruitment rates fluctuate linearly with the environment and are independent on other individuals' fitness; \textit{ii)} variability in mortality, in which environment affects mortality instead of recruitment rates; \textit{iii)} the case in which recruitment rates are re-scaled by the local averaged fitness, with the idea that adjacent neighbors are simultaneously competing for the same place. Cases \textit{i)} and \text{ii)} lead to identical dynamics in which fitness varies linearly, so they have been analyzed as the same one. We have focused our study on computing mean coexistence times as a measure of stability, finding scaling laws as a function of both demographic and environmental noise parameters.

In the linear fitness case, environmental fluctuations always reduce the time of species coexistence. In the relative fitness case, instead, a more complex situation is observed, and the correlation of the environment plays a crucial role: coexistence times are reduced for slowly varying environments and much increased for rapidly changing environments. The latter case can be understood as a direct consequence of Chesson's storage effect \citep{Chesson2000}. In a fast changing environment, no species can be optimal under all environmental conditions. The idea behind Chesson's mechanism is that species store the gains (increase in per-capita growth rate) achieved in those periods in which their fitness was higher, in order to face population losses in disadvantageous periods. It is worth noting that our results for large correlation time of the environment are congruent with the findings of \citep{Chesson1981} when generations are non-overlapping (i.e. when all individuals in the community are renewed before changing the environment). In our model, generations are always overlapping (one individual is killed and replaced each $\Delta t=1/N$), but one can think that, when the environment changes very slowly, the effective overlap becomes zero between consecutive switches of the environment.

Our study offers a unifying framework able to describe a variety of phenomena reported in the literature on the impact of environmental noise in neutral communities. For instance, in \citep{Vergassola2015}, species growth rates fluctuate linearly with the environment, and in such a case variability reduces the time of coexistence independently of the correlation of the environment \citep{Vergassola2015}. This observation is consistent with the first case studied here. On the other hand, another recent work \citep{Nadav2016} analyzes the spatially-explicit version of the voter-model with speciation, in which species fitness is modulated as the  the relative fitness case studied here, evidencing the storage effect and therefore the positive/negative impact of the environment on species richness depending on its correlation time \citep{Nadav2016}.

In our view, if environmental variability shapes recruitment rates, the relative fitness case constitutes the most realistic scenario from a biological point of view. Otherwise, if the environment shapes species mortality, the linear fitness case would be more appropriate in the modeling approach. However, further experimental analysis should be done to clarify this point. According to that, a final question arises: given a set of experimental data, what are the observables that help us to identify the dynamics that better describes our system? How to introduce endogenous variability in synthetic or natural communities so to promote species coexistence? This constitutes a fundamental question that should be carefully addressed when analyzing the impact of the environment in the organization of natural or experimental ecological communities.

Future perspectives would be to analyze more general situations in which population size is not constrained and both recruitment and mortality rates vary in time.
It may be of interest to consider correlations between time-varying recruitment and mortality.
In addition, we are interested in studying how environmental noise modulates species richness when neutrality is mildly broken, for instance by including preferred habitats for some species \citep{Borile2015,Pigolotti-Cencini2010} or by means of some degree of intraspecific competition \citep{Borile2012, Pigolotti-Cencini2013}. All these ingredients have been shown to enhance species coexistence. Finally, it would be interesting to introduce mutualistic/antagonistic ecological interactions among species in the community \citep{Allesina2012,Suweis2013}, leading to a complex dependency on the environment, as interactions among species would entangle all species fitness.

\section{Acknowledgments}
We thank Nadav Shnerb for his feedback on this work, as well as for suggesting us the analytical solution provided in Appendix D. 
We also thank Massimo Vergassola and Miguel A Mu\~noz for useful discussions. A. M. and J.H. acknowledge the support of the University of Padova (PRAT2014-CPDA148037).

\appendix

\section{Master equation and Kramers-Moyal expansion}
\label{sec:appendix1}
In this appendix we derive an approximate equation, valid for large populations, for a community of $N$ individuals of species $A$ and $B$, following a voter-model like dynamics in which individual fitnesses may depend on environmental conditions and local species abundance. In a well-mixed scenario, the state of the system is fully represented by the number of individuals of species $A$, $n_A=0,...,N$, and the state of the environment, $\epsilon=\pm1$; species A abundance is represented by $x=n_A/N$. The dynamics can be mapped into the following set of ``chemical reactions'':

\begin{equation}
\begin{array}{ccc}
 A + B &\overset{\lambda_A(x,\epsilon)}{\longrightarrow}& A + A\\
 A + B &\overset{\lambda_B(x,\epsilon)}{\longrightarrow}& B + B,\\
\end{array}
\label{eq:reaction-general}
\end{equation}
where $\lambda_A$ and $\lambda_B$ represent, respectively, the rate at which species $A$ and $B$ colonize positions occupied by the other species, that we identify with their fitness, 
and in general may depend on the species A density, $x=n_A/N$ (note that $n_B/N=1-x$), and the state of the environment, encoded in the variable $\epsilon$.

We also refer to a general situation in which the environment can switch between multiple states (i.e. more than two), $\epsilon\rightarrow\epsilon'$, with constant transition rates $k(\epsilon\rightarrow\epsilon')$. The probability of finding the environment in a state $\epsilon$ at time $t$, $P(\epsilon,t)$, follows the Master equation:

\begin{equation}
 \partial_t P(\epsilon,t)=\sum_{\epsilon'}\Big[ k(\epsilon'\rightarrow\epsilon)P(\epsilon',t)-k(\epsilon\rightarrow\epsilon')P(\epsilon,t) \Big]
 \label{eq:master-equation-env}
\end{equation}

Similarly, the Master equation for the probability of finding the system in a state $(n_A,\epsilon)$ at time $t$ (that for simplicity we also call $P(n_A,\epsilon,t)$) is:
\begin{eqnarray}
\partial_t P(n_A, \epsilon,t) = (E_{n_A}^--1)\left[ \lambda_A(n_A/N,\epsilon) n_A \dfrac{N-n_A}{N} P(n_A,\epsilon,t)\right] +\nonumber\\
 (E_{n_A}^+-1)\left[\lambda_B(n_A/N,\epsilon) (N-n_A) \dfrac{n_A}{N} P(n_A, \epsilon,t)\right]+\nonumber\\
\sum_{\epsilon'}\Big[ k(\epsilon'\rightarrow\epsilon) P(n_A, \epsilon') - k(\epsilon\rightarrow\epsilon') P(n_A, \epsilon,t) \Big] \nonumber\\
\label{eq:master-equation}
\end{eqnarray}
where we have introduced the operators $E_{n_A}^\pm$: $E^\pm_{n_A} f(n_A,\epsilon)=f(n_A\pm1,\epsilon)$.

We can rewrite the equation in terms of the density of individuals of species $A$, $x=n_A/N$ (also called $P(x,\epsilon,t)$ for simplicity):
\begin{eqnarray}
\partial_t P(x, \epsilon,t) = (E_x^--1)\left[N \lambda_A(x,\epsilon) x(1-x) P(x,\epsilon,t)\right]+\nonumber\\
 (E_x^+-1)\left[N \lambda_B(x,\epsilon) x(1-x) P(x, \epsilon,t)\right] +\nonumber\\
\sum_{\epsilon'}\Big[ k(\epsilon'\rightarrow\epsilon) P(x, \epsilon') - k(\epsilon\rightarrow\epsilon') P(x, \epsilon,t) \Big],
\label{eq:master-equation-x}
\end{eqnarray}
where we have already skipped the $1/N$ term coming from the Jacobian, with the new operators $E_x^\pm$: $E_x^\pm f(x,\epsilon)=f(x\pm\dfrac{1}{N},\epsilon)$. Then we perform a Kramers-Moyal expansion \citep{Gardiner} in terms of the density $x=n_A/N$ keeping the state of the environment, $\epsilon$, fixed, introducing in Eq. \ref{eq:master-equation-x} the approximation $(E_x^\pm-1)\simeq \pm \dfrac{1}{N}\partial_x + \dfrac{1}{2N}\partial_x^2$. Finally, we obtain the following (pseudo) Fokker-Planck equation:
\begin{eqnarray}
 \partial_t P(x, \epsilon, t) &= -\partial_x\left[(\lambda_A(x,\epsilon)-\lambda_B(x,\epsilon))x(1-x)P(x,\epsilon,t)\right]+\nonumber\\
&  \dfrac{1}{2N}\partial_x^2 \left[(\lambda_A(x,\epsilon)+\lambda_B(x,\epsilon))x(1-x) P(x,\epsilon,t)\right] +\nonumber\\
&  \sum_{\epsilon'}\Big[ k(\epsilon'\rightarrow\epsilon) P(x, \epsilon') - k(\epsilon\rightarrow\epsilon') P(x, \epsilon,t) \Big].
 \label{eq:fokker-planck}
\end{eqnarray}

Eq. \ref{eq:fokker-planck} can be mapped into the (Ito) Langevin equation:
\begin{eqnarray}
 \dot x &=& (\lambda_A(x,\epsilon)-\lambda_B(x,\epsilon))x(1-x) +\nonumber\\
 && \dfrac{1}{\sqrt{N}}\sqrt{ (\lambda_A(x,\epsilon)+\lambda_B(x,\epsilon)) x(1-x)} \xi(t),
 \label{eq:langevin-voter}
\end{eqnarray}
with $\xi(t)$ a zero-mean Gaussian white noise of unit amplitude, $\langle \xi(t)\xi(t') \rangle = \delta(t-t')$, and in which the environmental variable $\epsilon=\epsilon(t)$ follows its independent dynamics described by Eq. \ref{eq:master-equation-env}. Although the previous equivalence may result intuitive if the reader has some background on the equivalence between Fokker-Planck and Langevin descriptions, it is explicitly shown in the following section.
\subsection{Langevin equation with environmental noise}
\label{sec:langevin}
For convenience, we start from a generic (Ito) Langevin equation using the differential notation \citep{Gardiner}:

\begin{equation}
 d x = a(x,\epsilon,t)dt + b(x,\epsilon,t)dW,
 \label{eq:langevin-dw}
\end{equation}
where $dW = \xi dt$ and the dynamics of $\epsilon$ is described by Eq. \ref{eq:master-equation-env}. Given a generic function $f(\epsilon)$, using Eq. \ref{eq:master-equation-env} we can write its differential form $df(\epsilon)$:
\begin{eqnarray}
d \langle f \rangle  &=& \sum_{\epsilon} f(\epsilon) \partial_t P(\epsilon,t) dt \nonumber\\
&=& \sum_{\epsilon,\epsilon'} f(\epsilon) \Big( k(\epsilon'\rightarrow\epsilon) P(\epsilon',t)-k(\epsilon\rightarrow\epsilon') P(\epsilon,t) \Big) dt \nonumber\\
&=&\sum_{\epsilon} P(\epsilon,t) \sum_{\epsilon'}\Big( f(\epsilon') - f(\epsilon) \Big) k(\epsilon\rightarrow\epsilon') dt \nonumber\\
&=& \sum_{\epsilon} P(\epsilon,t) df(\epsilon).
\end{eqnarray}
Similarly, given a generic function $F(x,\epsilon)$, one can write its differential form:
\begin{eqnarray}
dF(x,\epsilon) = \partial_x F(x,\epsilon)dx + \dfrac{1}{2}\partial_x^2 F(x,\epsilon) dx^2 + O(dx^3) \nonumber\\
+ \sum_{\epsilon'}\Big[ F(x,\epsilon')-F(x,\epsilon) \Big] k(\epsilon\rightarrow\epsilon') dt,
\end{eqnarray}
whose expected value is:
\begin{eqnarray}
d \langle F \rangle &=& \sum_{\epsilon} \Big\langle \partial_x F(x,\epsilon) \left(a(x,\epsilon,t) dt + b(x,\epsilon,t) dW\right) +\nonumber\\
      && \dfrac{1}{2}\partial_x^2 F(x,\epsilon) \left( a(x,\epsilon,t)^2 dt^2  + \right.\nonumber\\
      && \left. 2 a(x,\epsilon,t) b(x,\epsilon,t) dt dW + b(x,\epsilon,t)^2 dW^2 \right)+\nonumber\\
 &&  \sum_{\epsilon'}\Big[ F(x,\epsilon')-F(x,\epsilon) \Big] k(\epsilon'\rightarrow\epsilon) dt \Big\rangle P(\epsilon,t) \nonumber\\
 &=& \sum_{\epsilon}\Big\langle a(x,\epsilon,t) \partial_x F(x,\epsilon) + \dfrac{1}{2}b(x,\epsilon,t)^2 \partial_x^2 F(x,\epsilon) +\nonumber\\
 && \sum_{\epsilon'}\Big[ F(x,\epsilon') - F(x,\epsilon) \Big] k(\epsilon'\rightarrow\epsilon) \Big\rangle P(\epsilon,t) dt + O(dt^2),\nonumber\\
\end{eqnarray}
where in the last step we have used that $\langle \partial_x^2 F(x,\epsilon) b(x,\epsilon,t)^2 dW^2\rangle=\langle \partial_x^2 F(x,\epsilon) b(x,\epsilon,t)^2 \rangle dt$ and that $\langle \partial_x F(x,\epsilon) b(x,\epsilon,t) dW\rangle = 0$ and $\langle \partial_x^2 F(x,\epsilon) a(x,\epsilon,t) b(x,\epsilon,t) dW\rangle = 0$ \citep{Gardiner}.

From this equation, we can write the time derivative of the expected value of $F$ in terms of the probability distribution $P(x,\epsilon,t)$:
\begin{eqnarray}
\dfrac{d}{dt}\langle F \rangle &=
 \int dx \sum_{\epsilon} \left( a(x,\epsilon,t) \partial_x F(x,\epsilon) + \dfrac{1}{2}b(x,\epsilon,t)^2 \partial_x^2 F(x,\epsilon) \right. + \nonumber\\
& \left. \sum_{\epsilon'}\Big[ F(x,\epsilon') - F(x,\epsilon) \Big] k(\epsilon\rightarrow\epsilon') \right) P(x,\epsilon,t) \nonumber\\
&= \int dx \sum_{\epsilon} F(x,\epsilon) \left(
 -\partial_x\left[ a(x,\epsilon,t) P(x,\epsilon,t) \right] \right. + \nonumber\\
 &\left. \dfrac{1}{2}\partial_x^2\left[b(x,\epsilon,t)^2 P(x,\epsilon,t) \right] 
 \right.\nonumber\\
& \left. +\sum_{\epsilon'}\Big[ k(\epsilon'\rightarrow\epsilon) P(x,\epsilon',t) - k(\epsilon\rightarrow\epsilon') P(x,\epsilon,t) \Big]\right).\nonumber\\
 \label{eq:almostfp1}
\end{eqnarray}
On the other hand, we can also write:

\begin{equation}
\dfrac{d}{dt}\langle F \rangle = \int dx \sum_{\epsilon} F(x,\epsilon) \partial_t P(x,\epsilon,t).
 \label{eq:almostfp2}
\end{equation}
Eq. \ref{eq:almostfp1} and \ref{eq:almostfp2} should be equivalent for any choice of $F(x,\epsilon)$, so we finally find that the (Ito) Langevin equation, Eq. \ref{eq:langevin-dw}, in which the dynamics of the environmental variable is described by the Master equation, Eq. \ref{eq:master-equation-env}, is equivalent to the following (pseudo) Fokker-Planck equation:
\begin{eqnarray}
 \partial_t P(x,\epsilon,t) = -\partial_x\left[ a(x,\epsilon,t) P(x,\epsilon,t) \right] +\nonumber\\
 \dfrac{1}{2} \partial_x^2 \left[b(x,\epsilon,t)^2 P(x,\epsilon,t)\right] + \nonumber\\
 \sum_{\epsilon'}\Big[ k(\epsilon'\rightarrow\epsilon) P(x,\epsilon',t) - k(\epsilon\rightarrow\epsilon') P(x,\epsilon,t) \Big].
\end{eqnarray}
In the case of Eq. \ref{eq:fokker-planck}, we obtain that it is equivalent to Eq. \ref{eq:langevin-voter} with an independent dynamics for the environment, Eq. \ref{eq:master-equation-env}.

\section{Stationary distribution with DMN}
\label{sec:appendix2}
In this appendix we briefly introduce the main equations to describe a stochastic process driven by DMN.
For a general and detailed review we refer to \citep{Bena2006}.

To gain some generality, we refer to a more general stochastic differential equation with DMN:

\begin{equation}
 \dot x = f(x) + \epsilon g(x),
 \label{eq:pdmp}
\end{equation}
in which the environmental variable $\epsilon$ alternates between the states $\epsilon=\pm1$ at rate $k$. With this choice, noise has zero mean, unit variance, and correlation time $\tau=(2k)^{-1}$. Let us notice that any equation of the form $\dot x = h(x,\epsilon)$ can be rewritten as Eq. \ref{eq:pdmp} by taking $f(x)=\dfrac{h(x,+1)+h(x,-1)}{2}$ and $g(x)=\dfrac{h(x,+1)-h(x,-1)}{2}$.

The temporal evolution of the marginal probability density $P(x,t)=P(x,+1,t)+P(x,-1,t)$ is given by \citep{Bena2006}:
\begin{eqnarray}
 \partial_t P(x,t) &=& - \partial_{x} \left[f(x) P(x,t) \right] +\nonumber\\
 &&\partial_{x} g(x) \int_{-\infty}^t dt' \exp\Bigg[-\Bigg(f'(x) +\nonumber\\
 &&f(x)\partial_x - \dfrac{1}{\tau}\Bigg)(t-t')\Bigg]\partial_x\left[g(x)P(x,t')\right] \nonumber\\
 &\equiv &\mathcal{L} P(x,t),
\end{eqnarray}
where the correlation time is $\tau=(2k)^{-1}$. Although  $\mathcal{L}$ is an intricate integro-differential operator, the explicit form of the stationary solution can be found  under rather general conditions \citep{Bena2006}, obtaining:

\begin{equation}
 P_{st}(x)= C \dfrac{g(x)}{D(x)} \exp\left[\dfrac{1}{\tau} \int_x \dfrac{f(x')}{D(x')}dx'\right],
 \label{eq:pst}
\end{equation}
where $C$ is a normalization constant and where we have introduced the effective diffusion coefficient

\begin{equation}
  D(x) = - (f(x)+g(x))(f(x)-g(x)).
  \label{eq:D}
\end{equation}

In the first case studied in the main text, Eq. \ref{eq:langevin-voter-env-I} of the manuscript, $f(x)=0$ and $g(x)=\sigma x (1-x)$. If one tries to apply the formula for the stationary distribution, the argument for the exponential term vanishes and its simply becomes $P_{st}(x) \propto g(x)^{-1}$. However, such a distribution cannot be normalized in the interval $[0,1]$, which tell us that there is not a stationary distribution for the process described by Eq. \ref{eq:langevin-voter-env-I}.

In the second case studied in the paper, corresponding to Eq. \ref{eq:langevin-voter-env-II}, $g(x)=\dfrac{2\sigma x(1-x)}{1-\sigma^2(2x-1)^2}$ and $f(x)=-\sigma(2x-1)g(x)$. Introducing these elements in Eq. \ref{eq:pst} and integrating over $[0,1]$ to fix the normalization constant, we obtain Eq. \ref{eq:pst-dmn} of the main text.

\section{Mean first passage time with DMN}
\label{sec:appendix3}
In this appendix we briefly review the main equations to find mean-first passage times of a stochastic process with DMN. Following the work of Sancho \citep{Sancho}, the analytical approach consists in calculating the ``backwards''  equation for $P(x,t)$ (with initial condition $P(x,0)=\delta(x-x_0)$ and $\epsilon(0)=-1$), so that the backwards dynamics can be represented by $\partial_t P(x,t|x_0,t_0) = L^\dag_{x_0} P(x,t|x_0,t_0)$, with $L^\dag_{x_0}$ the backwards evolution operator. Then, the mean-first passage time starting from $x_0$, $T(x_0)$, can be obtained by solving the equation $L^\dag_{x_0} T(x_0)=-1$ (given the boundary conditions). The explicit form of $L^{\dag}_{x_0}$ for the case of DMN can be explicitly computed \citep{Sancho}, obtaining the following differential equation for $T(x_0)$:
\begin{eqnarray}
&  D(x_0) T''(x_0) + \Bigg(\dfrac{1}{\tau} f(x_0) -\nonumber\\
&(f(x_0)+ g(x_0))(f'(x_0)-g'(x_0))\Bigg)T'(x_0)+\dfrac{1}{\tau}=0,\nonumber\\
  \label{eq:mean-first-passage-time}
\end{eqnarray}
where the functions $f$, $g$ and $D$ have been defined in Appendix B.
Eq. \ref{eq:mean-first-passage-time} has to be solved with the boundary conditions of the particular problem; for instance, we impose that $T(1/N)=T(1-1/N)=0$ for (artificial) absorbing barriers at $x=1/N$ and $x=1-1/N$.

In the first case studied in the paper, i.e. the process described by Eq. \ref{eq:langevin-voter-env-I}, we obtain Eq. \ref{eq:mfpt1} of the main text, which can be solved analytically, and the solution corresponds to Eq. \ref{eq:mfpt1-solution}. In most other cases, however, Eq. \ref{eq:mean-first-passage-time} cannot be solved exactly, but some information can be retrieved using approximate methods. In the work of Sancho \citep{Sancho}, he studied the case in which the process escapes from a fix point $x_0$ (so that $f(x_0)=0$), overcoming a barrier located at $x^*$, obtaining a simple modified Arrhenius's law for the mean-first passage time:

\begin{equation}
 T_N(x_0)=\dfrac{2\pi}{|f'(x_0)f'(x^*)|} \exp\left(U(x^*)-U(x_0)\right)
 \label{eq:arrhenius}
\end{equation}
in which the Arrhenius factor is given by:

\begin{equation}
U(x) = -\dfrac{1}{\tau} \int^{x} \dfrac{f(x')}{D(x')}dx'.
\end{equation}
We can use this approximate result for the second case studied in the main text, Eq. \ref{eq:langevin-voter-env-II}, obtaining the leading scaling relation of Eq. \ref{eq:ext2}.
However, although this procedure provides a valid scaling relation for this case, it has not been rigorously derived. A more formal derivation from Eq. \ref{eq:mean-first-passage-time} can be found in Appendix \ref{sec:appendix4}.

\section{Derivation of Eq. \ref{eq:ext2}}
\label{sec:appendix4}
In this Appendix we present the calculation of the scaling relation for the second case studied in the paper, Eq. \ref{eq:ext2}.
Eq. \ref{eq:mean-first-passage-time} can be written in terms of $Q(x_0)=T'(x_0)$, which in our case obeys the condition $Q(1/2)=0$ for symmetry reasons.
Introducing the corresponding expressions of $f(x)$, $g(x)$ and $D(x)$ (see Appendix B) in Eq. \ref{eq:mean-first-passage-time}, one can find an exact expression of $Q(x)$:
\begin{eqnarray}
Q(x_0) &=& \dfrac{1}{4\tau\sigma^2} \dfrac{1- \sigma (2x_0-1)}{(x_0(1-x_0))^{1+\frac{1}{2 \tau }}}  \times \\ 
\nonumber & &\left[(\sigma +1) B_{\frac{1}{2}}\left(\frac{1}{2 \tau },\frac{1}{2 \tau }\right)-2 \sigma  B_{\frac{1}{2}}\left(\frac{1}{2 \tau },1+\frac{1}{2 \tau }\right) \right.\\
\nonumber & & \left.-(\sigma +1) B_{x_0}\left(\frac{1}{2 \tau },\frac{1}{2 \tau }\right)+2 \sigma  B_{x_0}\left(\frac{1}{2 \tau },1+\frac{1}{2 \tau }\right)\right],
\end{eqnarray}
where $B_z(a,b)$ is the incomplete Beta function. The mean-extinction time is given by $T(x_0)=\int_{x_0}^{x^*=1/N} dx Q(x)$. As we are interested in the asymptotic behavior for large $N$, we can just look at the contribution of the divergence given by the upper limit of the integral for $x^*\rightarrow0$. In this regime, the contribution of the third and fourth Beta functions vanishes, and $Q(x\rightarrow0) \propto x^{-1-1/2\tau}$, that, when integrated and evaluated at $x=1/N$, leads to Eq. \ref{eq:ext2} of the main text.

\bibliographystyle{elsarticle-harv} 

\end{document}